\documentclass{article}                
\usepackage{graphicx,epsf,subfigure,pstricks,pst-node,psfrag,amsthm,amssymb,amsmath,url,inputenc}
\usepackage{float}
 
\begin{document}
\title{\textbf{\huge EPA Particulate Matter Data - Analyses using Local Control Strategy}}

\author{Robert L. Obenchain and S. Stanley Young}
\date{December 2022}
\maketitle

\begin{abstract}%
\textit{Statistical Learning} methodology for analysis of large collections of  cross-sectional \textit{observational data} can be
most effective when the approach used is both \textit{Nonparametric} and \textit{Unsupervised}. We illustrate use of our
\textit{NU Learning} approach on 2016 US environmental epidemiology data that we have made freely available. We encourage other
researchers to download these data, apply whatever methodology they wish, and contribute to development of a broad-based ``consensus
view'' of potential effects of Secondary Organic Aerosols (volatile organic compounds of predominantly biogenic or anthropogenic
origin) within PM$_{2.5}$ particulate matter on circulatory and/or respiratory mortality. Our analyses here focus on the question:
``Are regions with relatively high air-borne biogenic particulate matter also expected to have relatively high circulatory
and/or respiratory mortality?''
\end{abstract}

\noindent Keywords: Nonparametric Unsupervised Learning; Local Control Strategy; Clustering as Matching; Permutation Distributions;
Random Forests and Partial Dependence Plots.


\section{Introduction} 

When a researcher downloads a large file of cross-sectional data from the internet, it’s too late to worry about potential
deficiencies in the \textit{experimental design} characteristics of observational data. One could discard some parts of the data
to make the remainder ``look better'': e.g. look more \textit{blocked} and/or \textit{balanced}. But using \textit{much less than
all of the available data} opens a researcher up to potential accusations of ``cherry-picking'' or having ``gardened'' the data
to get some desired result, Glaeser (2006).

Here we use R-functions from CRAN-packages, R Core Team (2022), that implement a highly adaptive strategy that can robustly
``\textit{design an analysis}'' of potentially confounded variables. We apply this approach to the data.frame, \textit{pmdata},
that is part of the newest Version ($1.4$) of the \textit{LocalControlStrategy} R-package, Obenchain (2015-2022). This data.frame
contains a total of $122$ variables that quantify diverse characteristics of $2,980$ individual US Counties. To form \textit{pmdata},
we merged EPA data downloaded from \url{https://doi.org/10.5281/zenodo.5713903} with data from \textit{CDC Wonder} (2017),
using \textit{fips} codes. Our PDF file of R-documentation for \textit{pmdata}, which can be downloaded from
\url{https://CRAN.R-project.org/package=LocalControlStrategy}, provides ``shortened'' names for all $122$ variables as well as any
original EPA names and variable descriptions.

Researchers wishing to perform analyses using software other than R can access the CSV data file we have posted on \textit{dryad},
Young and Obenchain (2022), \url{https://doi.org/10.5061/dryad.63xsj3v58}.  This CSV file contains only the $25$ variables that we initially considered using in our analyses of the $2,973$ US Counties without any \textit{missing values} for the $11$ variables
listed in Table 1.
 
Since we accessed \textit{CDC Wonder} data on April 19, 2022 (more than 19 months after EPA researchers), our ``2016'' CDC data
may not actually be identical to that analyzed in PYE et al. (2021). A more important difference is that our analyses here will
focus on CDC ``Age Adjusted'' rates of Circulatory and/or Respiratory mortality, \textit{AACRmort}, while published EPA analyses
used only CDC ``crude'' (or raw) mortality rates.

Here, we illustrate use of the LC Strategy R-package to implement the \textit{NU Learning} approach of Obenchain (2015-2022). Its
three key-characteristics are:

[a] Users must specify both a single $y-$outcome variable and a single primary (potentially causal) variable that represents
either (1) a binary ``treatment choice'', $t=0$ or $t=1$, or else (2) a continuous primary ``exposure'' measure, $e$. The analyses
illustrated here use $y = AACRmort$ and $e = Bvoc$ (biogenic volatile organic compounds), which are two of the eleven variables
described in Table $1$.

[b] LC Strategy then starts by forming Clusters of observational units (here, US Counties) that are \textit{relatively well-matched} in an $X-$covariate space also specified by the analyst. Software supporting application of LC Strategy needs to provide statistical measures and graphics that help users make well-informed choices among both clustering algorithms and the best number, $K$, of clusters to use with each algorithm.  

[c] The primary output from LC Strategy for each clustering algorithm consists of two new variables: a cluster
ID value for each unit, and a local (within cluster) ``effect-size'' measure for each unit. In our final ``Reveal Phase'' of
LC Strategy, we will illustrate use of Local Rank Correlations (LRCs) between $AACRmort$ and $Bvoc$ within $K = 50$ Clusters
formed using the \textit{ward.D} algorithm.

\begin{figure}[!htbp]
\center{\includegraphics[width=4.5in]{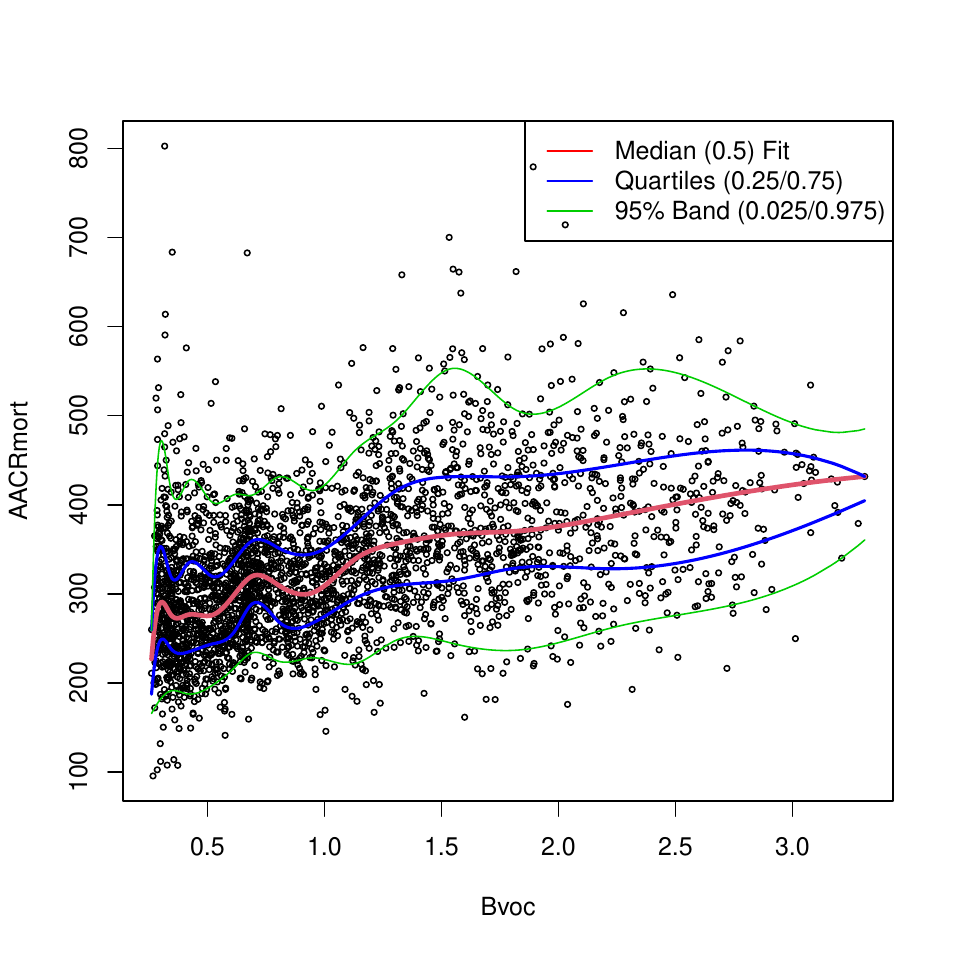}}
\caption{The above scatter for 2,973 US Counties within the contiguous 48 States plus Washington DC reveals
considerable variability in $y = AACRmort$ for any given value of $e = Bvoc$. The five (red, blue or green) regression
quantiles fitted using the quantreg $R-$package, Koenker (2005-2022), display considerable non-linearity. Thus, given only
the Bvoc level of an otherwise unspecified US County, it would clearly be unrealistic to expect a meaningful point-prediction
of its AACRmort count to result.}
\label{Fig01}  
\end{figure}

A good place to start our analyses is provided by Figure~\ref{Fig01}. This scatter plot displays $e = Bvoc$ ($\mu{g}/m^3$) from EPA models
using satellite data (on the horizontal axis) verses $y = AACRmort$ estimates from the CDC for $2,973$ US Counties. The $Bvoc$
component of PM$_{2.5}$ consists of \textit{volatile organic compounds} that are primarily biogenic (rather than anthropogenic) in
origin within ``Secondary Organic Aerosols'', Pye et al. (2021). $AACRmort$ values are ``Age Adjusted Circulatory and/or Respiratory mortality''
counts of deaths per 100,000 residents.

In natural products research, it is standard practice to examine fractions of raw natural material to figure out where
an ``active agent'' is located. There typically are many fewer compounds in the active fraction, so chemical identification
becomes more feasible. Starting in 1993, evidence pointing to particulate matter smaller than $2.5$ microns, thereafter
known as PM$_{2.5}$, became the proposed active fraction in the 1952 London Fog death event, Logan (1953).

In reality, PM$_{2.5}$ is a complex mixture of over four thousand chemical compounds which vary over both time and space
(urban vs. rural, etc.) PYE et al. (2021) appear to have followed a natural products strategy by arguing that ``Secondary Organic Aerosols'' (SOAs) could be ``the'' active agents in PM$_{2.5}$. Unfortunately, in the almost 30 years since PM$_{2.5}$ publications sponsored by the US EPA first appeared, neither measures of PM$_{2.5}$ itself, nor any fraction of it, has been convincingly shown
to be genuinely predictive of mortality at the level of individual US Counties and Cities.

Young, Kindzierski and Randall (2021)  suggest that the current EPA paradigm that ``PM$_{2.5}$ is, by itself, lethal'' is
wrong. While the efforts of Pye et al. (2021) may be noble, potential challenges to their proposals abound.

\section{Predicted rather than Observed Measures of SOAs} 

Although actual measurements of SOA components in air samples are apparently not currently feasible, various ``models'' for predicting their presence from measurements of other pollutants have apparently been used since Volkamer et al. (2006). The analyses presented in
Pye et al. (2021) appear to be based upon the EPA's \textit{Community Multiscale Air Quality} (CMAQ) System Models, U.S. EPA Office of Research and Development (2019). Unfortunately, neither the assumptions nor any limitations of those models are discussed in Pye et
al. (2021).

For example, a potential issue is that CMAQ may be loosely based upon \textit{Generalized Additive Model} (GAM) concepts. While
Pye et al. (2021) do reference Wood (2003, 2004 and 2022), their introduction claims that they use ``multiple linear regression'',
a methodology that makes strong and seemingly unrealistic assumptions for observational data. Since GAMs are a composite of several individual models, some sub-models could be fit in unspecified ``non-standard'' ways. All that we know as outside observers is that
these EPA models do a much better job of predicting CDC measures of Circulatory and/or Respiratory mortality (raw or age-adjusted)
than we would have expected. In short, we would much rather have had \textit{actual observed measurements of secondary organic aerosols from validated scientific instruments} to analyze here ...and to make available to other researchers.

\section{Variables of Primary Interest within CDC and EPA Data} 

There are $25$ variables (columns) and $2,973$ observations (rows) in the ``AnalysisFile.csv'' file that we have archived at \textit{dryad}, Young and Obenchain (2022), while the analyses presented here actually use only the $11-$variables described in Table $1$.
Six US Counties out of the original $2,980$ had NA codes on the rather important PREMdeath variable (fips = 8033, 30055, 31091, 46069, 46073 and 49031), while a $7^{th}$ NA code occurred for fips = 32011 in the IncomIEQ variable, which turned out to be a relatively unimportant confounder.

The analyses presented here will ultimately focus on ``potential causes'' of Circulatory and Respiratory Mortality within
individual US Counties. Our focus on the AACRmort variable from the CDC (rather than CRmort) stems primarily from our
previous experience [Obenchain, Young and Krstic (2019)] where the \textit{proportion of county residents over 65} was a key
predictor of local mortality. While we did perform some preliminary analyses using CRmort as the primary $y-$Outcome, we found
them to be both more complex and less self-consistent than the corresponding AACRmort analyses that we present here.

Our somewhat surprising and unexpected findings concerning distinct components of pmTOT (i.e. PM$_{2.5}$) focus
primarily upon Bvoc, although the Avoc and pmSO4 levels from the EPA are among the confounders considered in our final analyses.

\vspace{0.5cm}
\begin{tabular}{@{}lll@{}}
\multicolumn{3}{c}{\textbf{TABLE 1 -- Variable Information}} \\
\textbf{Name} & \textbf{Description} & \textbf{Range} \\
fips & Federal Information Processing code ($4$ or $5$ digits) & 1001-56045 \\
CRmort & Circulatory/Respiratory (crude) mortality (per 100K) & 64.8-1564 \\
AACRmort & Age Adjusted Circulatory/Respiratory mortality (per 100K) & 95.5-802.6 \\
pmTOT & Observed $PM_{2.5}$ level ($\mu$g/m$^3$) & 2.06-14.32 \\
Bvoc & Biogenic volatile organic compounds in pmTOT & 0.26-3.31 \\
Avoc & Anthropogenic volatile organic compounds in pmTOT & 0.23-2.89 \\
pmSO4 & Sulfate compounds in pmTOT  & 0.39-1.62 \\
ASmoke & Adult Smoking fraction & 0.007-0.412 \\
PREMdeath &  Premature Death rate & 0.03-0.66 \\		   
ChildPOV  &  Children living in Poverty (per 100K) & 2853-36469 \\
IncomIEQ  &  Income In-Equality rating & 2.9-8.9 \\
\end{tabular}
\vspace{0.5cm}

\begin{figure}[!htbp]
\center{\includegraphics[width=3.5in]{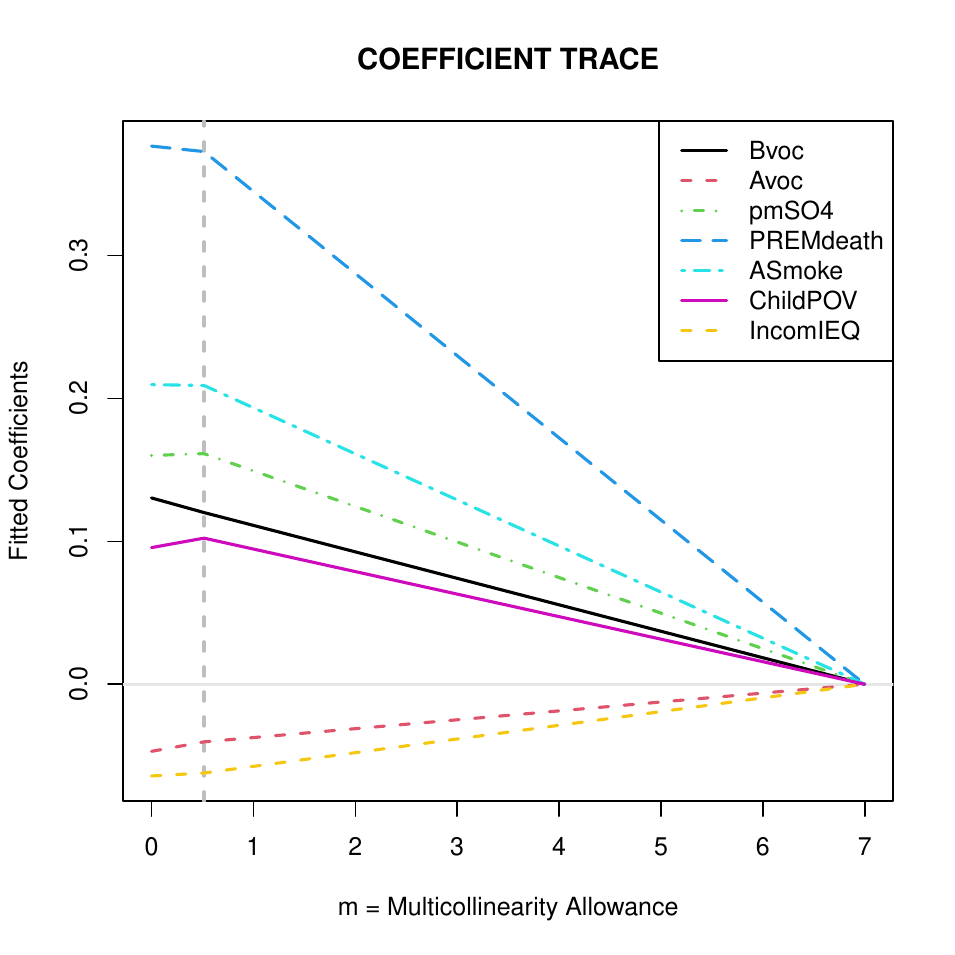}}
\caption{Following a modest initial extent of shrinkage to $m=0.51$, MSE-risk optimal relative-magnitudes for
$\beta-$coefficient estimates are determined, Obenchain (2022). In other words, overall ill-conditioning (redundancy) among these seven
$X-$variables appears minor. Since our main LC Strategy analyses will use AACRmort as the $y-$Outcome measure, we will
focus on Bvoc as our primary (potentially causal) Exposure variable. After all, the EPA Avoc and IncomIEQ variables have
only \textit{small and negative coefficient estimates}. The overall $R^2-$Coefficient of Determination for this linear model
is not particularly high (coincidentally, also $0.51$). Thus, true relationships here could be nonlinear. In fact, the ``local''
models developed here will be \textit{distribution-free} and correspond to much better fits.}
\label{Fig02} 
\end{figure}

Although we ultimately focus on \textit{Local} (within Cluster) models that are highly-flexible and statistically robust,  
we initially used \textit{multiple linear regression} to assess the ``overall'' extent of ill-conditioning (confounding) among
variables; see Figure~\ref{Fig02}. This \textit{Trace} display shows fitted regression coefficients on the \textit{Efficient
Generalized Ridge Regression} shrinkage path, Obenchain (2005-2022). This linear model attempts to predict $AACRmort$ using $Bvoc$,
$Avoc$, $pmSO4$, $PREMdeath$, $ASmoke$, $ChildPOV$ and $IncomIEQ$. While the order in which these seven ``right-hand-side''
variables are specified is irrelevant in classical linear regression, the first three are EPA environmental variables while the
last four (hopefully) quantify socioeconomic characteristics of county residents.

\newpage

\section{Applying Local Control Strategy} 

LC Strategy consists of Four Phases of analysis called, respectively : \textit{Aggregate, Confirm, Explore} and \textit{Reveal}.
Each analysis ``cycle'' through the available data typically uses a different Aggregation of observational units into a larger
and larger number, $K$, of smaller and smaller ``Clusters''. Potential results from each new cycle are ultimately \textit{Discarded}
unless they have been \textit{Confirmed} to provide \textit{Added Information} ...in the sense of being clearly different from
``purely random'' in ways outlined below.

The implicit statistical model used in LC Strategy is a simple ``one-way'' classification of within-cluster estimates. In the analyses
presented here, these estimates are ``Local Rank Correlations'' (Spearman LRCs) between an \textit{Exposure e-variable} and an \textit{Outcome y-variable}, both of which are potentially continuous measures ...i.e. have many more than just two levels.
Specifically, we discuss the special case where the $e-$variable is \textit{Bvoc} and the $y-$outcome measure is \textit{AACRmort}.
Thus, when the $j^{th}$ cluster contains $N_j$ US Counties, $N_j$ identical LRC estimates are contributed to the overall (across
cluster) distribution of LRC Effect-Sizes.

Note, in particular, that analyses focused on \textit{LRC estimates} implement a variation on the recommendation of Rubin (2008) for
analysts to be ``blinded'' to the numerical values of the \textit{y-outcome} ($AACRmort$) and primary \textit{exposure} ($Bvoc$)
variables. Specifically, the $LRC$ vector of values created via LC Strategy guides all subsequent primary analyses. Clusters with
$LRC$ estimates near zero suggest ``absence of any \textit{clear local relationship}'' between $Bvoc$ and $AACRmort$, while strongly
non-zero $LRCs$ suggest \textit{local relationships} that tend to be nearly monotone increasing or decreasing between $Bvoc$ and
$AACRmort$ \textit{within that particular cluster}.

While LC Strategy can be used by single researchers working alone, this strategy can also be used by a \textit{group of collaborating
researchers}. In fact, LC Strategy may be an ideal approach when the researches involved in a collaboration have ``odd combinations''
of qualifications ...such as: diverse experiences and unique perspectives on the overall issues involved. Use of LC Strategy plus
access to relevant data then makes it possible to seriously address the question: Can any shared, consensus position be reached? 

Our discussion of LC Strategy below will use the basic terminology outlined in Table $2$.

\vspace{0.5cm}
\begin{tabular}{@{}ll@{}}
\multicolumn{2}{c}{\textbf{TABLE 2 -- Basic LC Strategy Terminology}} \\
\textbf{Term} & \textbf{Description} \\ 
Observational Unit & An individual US County or Parish within the contiguous 48 States \\
                   & or the District of Columbia. \\
Cluster & A subgroup of Units that are relatively well-matched on their given \\
        & $X-$characteristics. \\
Cycle   & One ``pass'' through the data involving at least the Aggregate and \\
        & Explore phases of analysis. \\
Aggregate & The Phase of LC Strategy where a new set of Clusters are formed. \\
Confirm & The Phase where a given Aggregation can be shown to either be \\
        & ``Ignorable'' or to provide \textit{Added Information}. \\
Explore & The Phase where Aggregations are compared on Variance-Bias Trade \\
        & Offs and a decision to either Continue or Stop Cycling is made. \\
Reveal  & The final Phase of LC Strategy where the Distribution of Within \\
        & Cluster Local Rank Correlations is analyzed using randomForest() \\
        & functions and Partial Dependence plots. \\
\end{tabular}
\vspace{0.5cm}

\subsection{Forming a Hierarchical Clustering Tree}

\begin{figure}[!htbp]
\center{\includegraphics[width=4in]{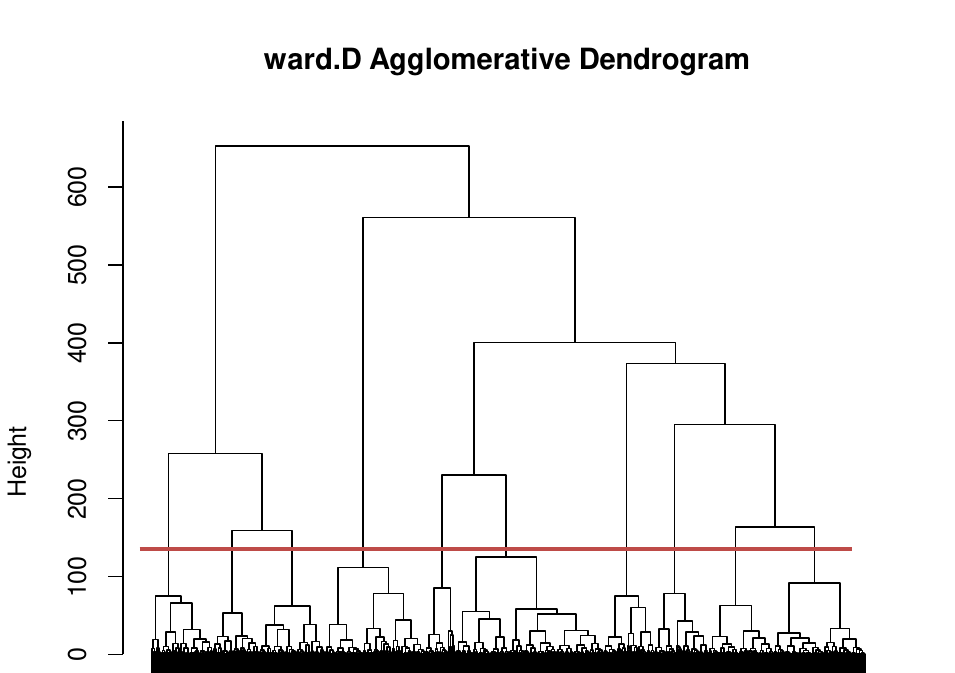}}
\caption{\label{Fig03} This Hierarchical Clustering \textit{Dendrogram} can be ``cut'' by a horizontal line
at any strictly positive Height. For example, note that the displayed red line produces 10 clusters.}
\end{figure}

The very first step in applying LC Strategy is to form a hierarchical clustering \textit{dendrogram}. The
LocalControlStrategy::LCcluster() function calls stats::prcomp() to calculate Mahalanobis distances using
standardized and orthogonal Principal Coordinates, Obenchain (1971) and Rubin (1980). These coordinates optimally quantify
``dissimilarity'' among US Counties using their possibly confounded $x-$characteristics. A dendrogram
(tree-like structure) can usually be ``cut'' by a horizontal line to produce any desired number of individual
clusters. Our example dendrogram, displayed in Figure~\ref{Fig03}, uses the default ``ward.D'' clustering algorithm. 

Either the divisive cluster::diana() method or any of six agglomerative alternatives (ward.D2, complete, average,
mcquitty, median, or centroid) could have been specified. However, ``single-linkage'' clustering is not
appropriate for use in LC Strategy because it can produce clusters that are more like ``strings'' than like compact
subgroups of most similar observational units.

\subsection{Confirming that a given Number of Clusters provide ``Added Information''}

Each new \textit{Aggregate Phase} of LC Strategy may be followed by a full or partial \textit{Confirm Phase}
analysis of the new distribution of within-cluster Local (Spearman) Rank Correlations (LRCs). Each new Clustering
provides \textit{Added Information} if, and only if, the \textit{empirical Cumulative Distribution Function} (eCDF) of
it's LRC-distribution is clearly different from the NULL eCDF that corresponds to \textit{Purely Random} assignments of
all observational units to the same number, ``K'', of Clusters of the exactly same size as the given clusters. This
NULL distribution and a two-sample Kolmogorov-Smirnov D-statistic can be efficiently computed (in about 5 seconds)
using Random Permutations. A numerically large D-statistic provides only \textit{partial confirmation}.

Unfortunately, the tabulated distribution of KS D-statistics assumes that both distributions are \textit{absolutely
continuous}, while LRC-distributions are clearly \textit{discrete}. Thus, LC Strategy uses a (computationally intensive)
\textit{permutation test}, Welch (1990), that is non-parametric. Simulating the $p-$value of an observed LRC
$D-$statistic uses the KSperm() function in the LocalControlStrategy $R-$package and typically requires an additional
75 seconds of computation to (potentially) achieve \textit{full statistical confirmation} ``adjusted for Ties''.

\begin{figure}[!htbp]
\center{\includegraphics[width=4.5in]{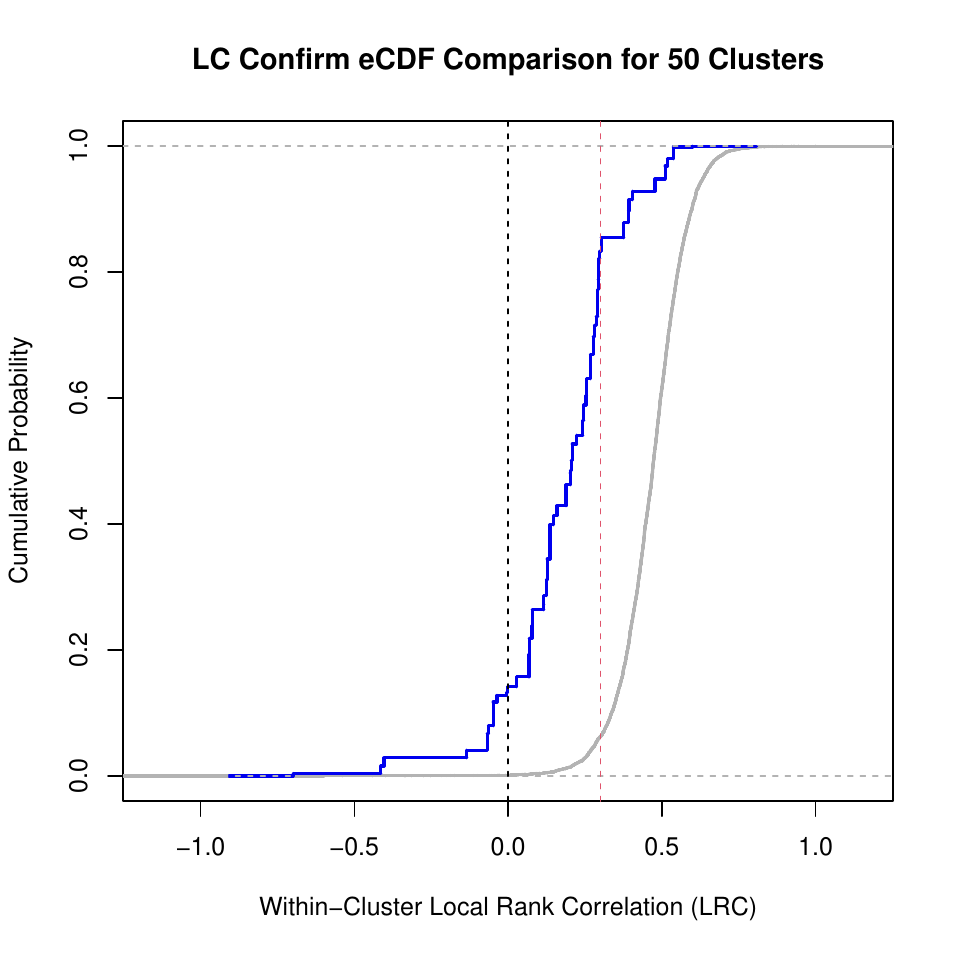}}
\caption{\label{Fig04} This plot displays a pair of empirical Cumulative Distribution Functions (eCDFs) and addresses
the question of whether they are statistically different. Note that the Kolmogorov-Smirnov D-statistic of $0.7835$
(maximum vertical separation between eCDFs) occurs here at $0.30$ on the horizontal axis ...marked by the dashed red line.}
\end{figure}

Figure~\ref{Fig04} displays the numerical example with $K = 50$ Clusters that we will use to illustrate \textit{confirm
phase testing}. We are ``jumping the gun'' here in the sense that we will argue next (in Subsection $4.3$) that $K = 50$ is
also the ``optimal'' number of Clusters to use in applying our LC Strategy for \textit{NU Learning}.
  
Note that the gray NULL eCDF appears to be rather \textit{smooth} in Figure~\ref{Fig04}. After all, it results from
pooling $100$ sets of LRC estimates from $50$ pseudo-clusters that were formed in a \textit{purely random and meaningless way}.
Thus the gray eCDF contains at most $5,000$ quite small steps. To truly provide ``added information'', the distribution underlying
the Blue eCDF from clusters of ``well matched'' US Counties must be clearly different from the purely random distribution
(from random pseudo-clusters) represented by the gray eCDF.

\begin{figure}[!htbp]
\center{\includegraphics[width=4in]{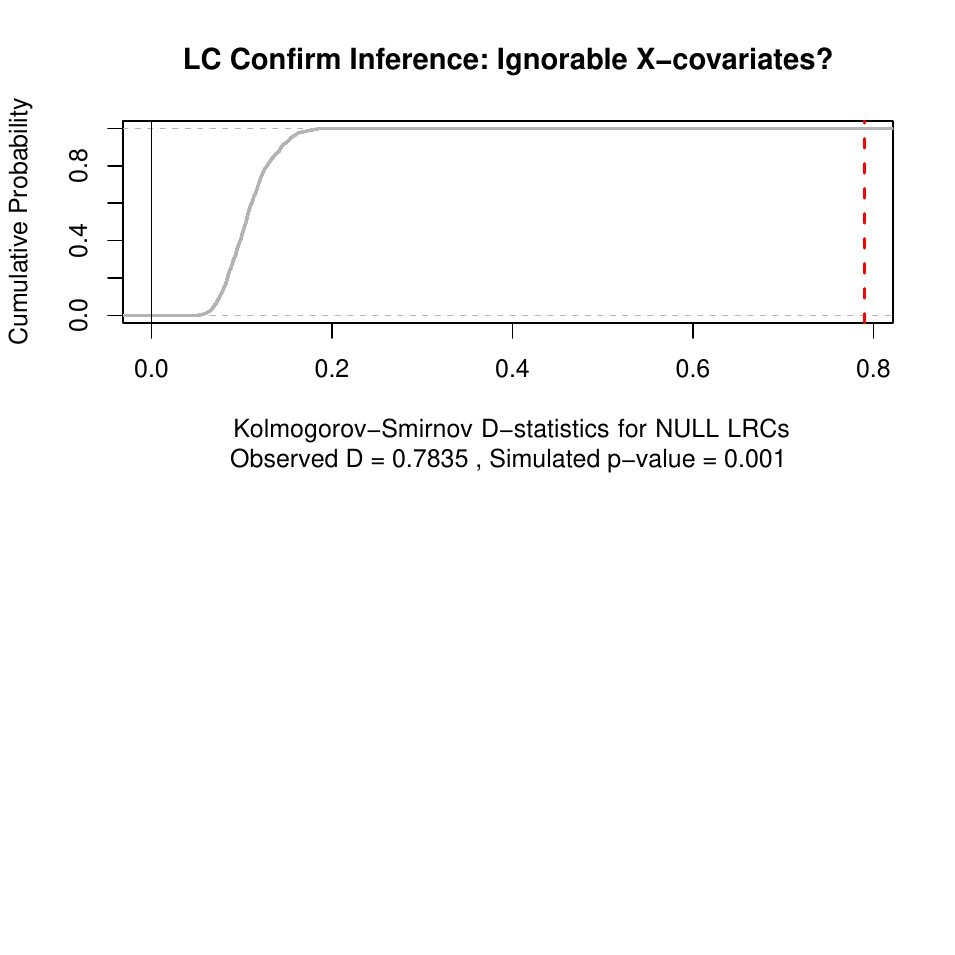}}
\caption{\label{Fig05} Since the traditional Kolmogorov-Smirnov test compares eCDFs from a pair of
\textit{Continuous Distributions}, it is severely biased when used to compare \textit{Discrete Distributions}.
Thus, simulation is used in LC Strategy to make valid comparisons between eCDF estimates. Since the largest observed NULL D-statistic
was only $0.237$, the Observed D of $0.7835$ almost surely has an actual $p-$Value of \textit{much less than}
$0.001$ (one in a thousand).} 
\end{figure}

Figure~\ref{Fig05} displays the NULL distribution where the \textit{largest observed NULL D-statistic} is only $0.237$.
This simulated NULL D-statistic is dwarfed by the Observed $D-$statistic of $0.7835$ in Figure~\ref{Fig04}. In other words,
the Blue empirical CDF variable, named ``LRC50'', provides abundant \textit{Added Information} in the sense that it is
clearly ``shifted to the left'' in Figure~\ref{Fig04} relative to the simulated gray NULL eCDF. The Blue eCDF is thus
\textit{Not Ignorable}! Furthermore, note that some Local Rank Correlation estimates from $K = 50$ clusters are
\textit{clearly negative}. 

\subsection{Choosing the Number of Clusters to Use in LC Strategy}

Especially when cross-sectional data are observational, the data typically contain isolated ``local'' effects as well
as overall (global) confounding from potentially highly correlated variables. The Nonparametric and Unsupervised
``pre-processing'' methods that characterize LC Strategy address these issues. Traditional methods based on multiple
regression are typically Global, Parametric and Supervised. They tend to make much stronger (and possibly unrealistic)
assumptions that yield questionable global predictions and potential extrapolations.

In sharp contrast, LC Strategy summarizes findings from several separate and flexible ``local'' models that are fit within \textit{clusters of ``most similar'' and/or relatively ``well-matched'' observational units}. The number of such local models
and separate clusters of observational units being used in a given Cycle is denoted by the symbol \textit{K}.

LC Strategy embraces many of the basic concepts and experiences outlined in Rubin (2008), Stuart (2010), van der Laan and
Rose (2010) and Stang et al. (2010). Published analyses using LC Strategy include: Obenchain and Young (2013); Lopiano, Obenchain and
Young(2014); Young, Smith and Lopiano (2017) and Obenchain, Young and Krstic (2019).

\begin{figure}[!htbp]
\center{\includegraphics[width=4.5in]{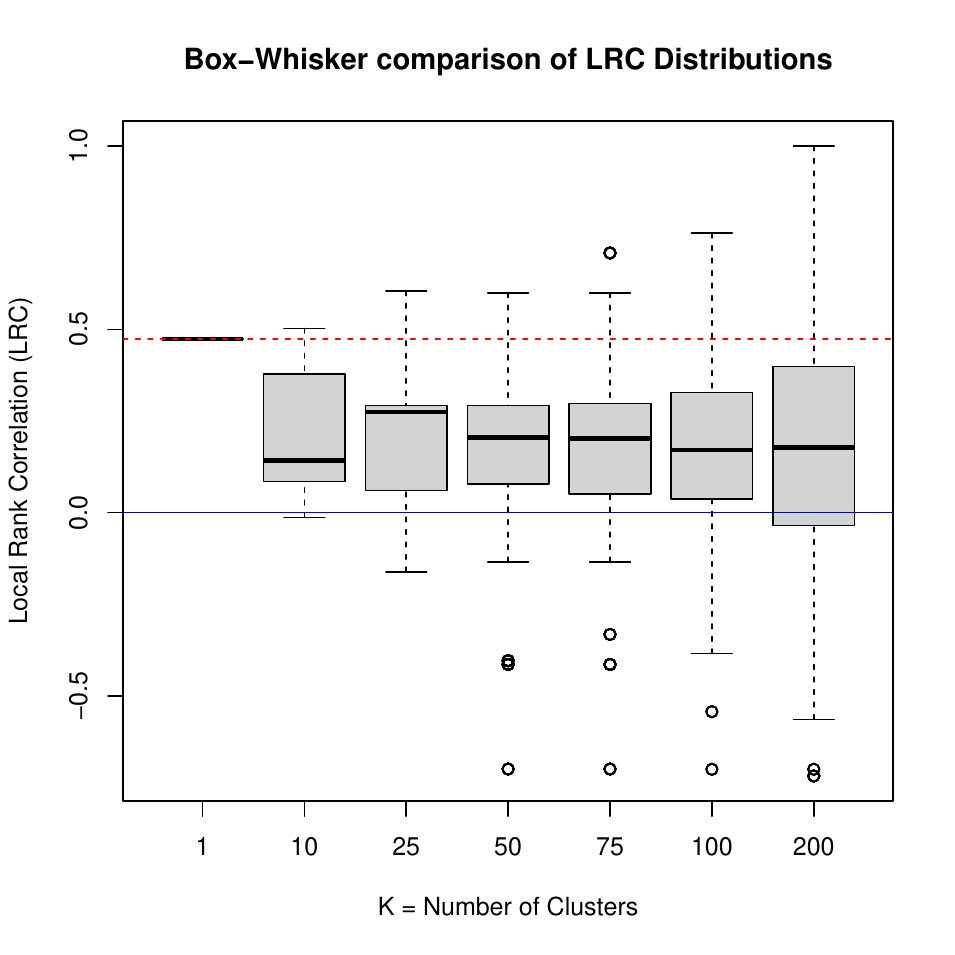}}
\caption{\label{Fig06} Here we see $7$ alternative Box and Whisker diagrams that show how the overall distribution of LRC
effect-sizes changes as the number of clusters, $K$, being formed increases from $1$ to $200$. Note that the LRC distribution
for $K = 50$ has the smallest interquartile range for $K > 1$ as well as shorter whiskers (both upwards and downwards) than
the LRC distributions for $K > 50$. Our recommended ``compromise'' choice is thus $K = 50$.} 
\end{figure}

Ultimately, we chose to use $K = 50$ Clusters of US Counties formed using only the six key $x-$Confounder variables
that made (potentially causal) ``common sense'' to us and repeatedly proved useful in many alternative LC analyses.
The resulting $50$ clusters represent mutually exclusive and exhaustive statistical ``Blocks'' of \textit{relatively
well-matched observational units}, and our resulting ``best'' Effect-Size distribution quantifies a potentially causal
relationship between the numeric $y-$Outcome variable AACRmort and our $e-$Exposure variable Bvoc.

\begin{figure}[!htbp]
\center{\includegraphics[width=5in]{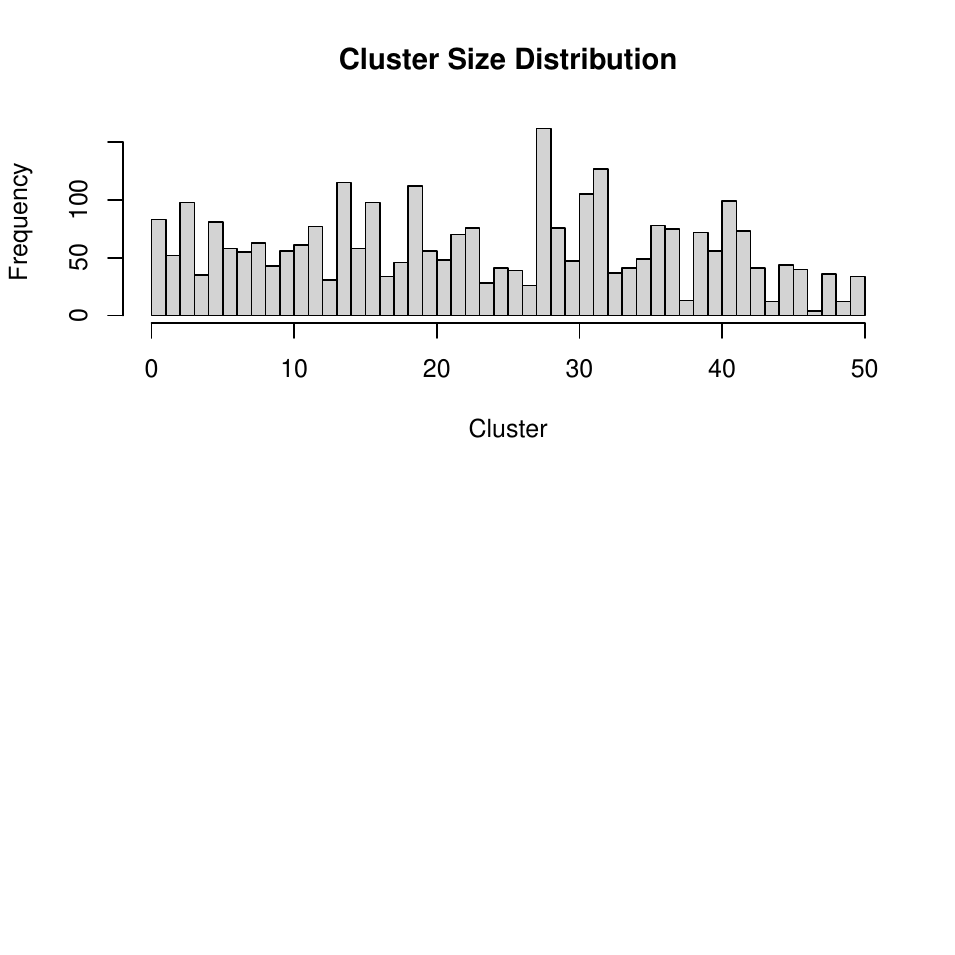}}
\caption{\label{Fig07} This histogram shows the number of US Counties within each of the 50 Clusters formed using observational data from the EPA, CDC and US Census. In a prospective study (e.g. a well ``designed'' experiment), all ``blocks'' of well-matched units are usually of nearly equal size. But that's an unlikely ``ideal'' situation when analyses must rely on observational data. The largest cluster contains 162 US Counties, while the smallest contains only four.}
\end{figure}

\begin{figure}[!htbp]
\center{\includegraphics[width=6.5in]{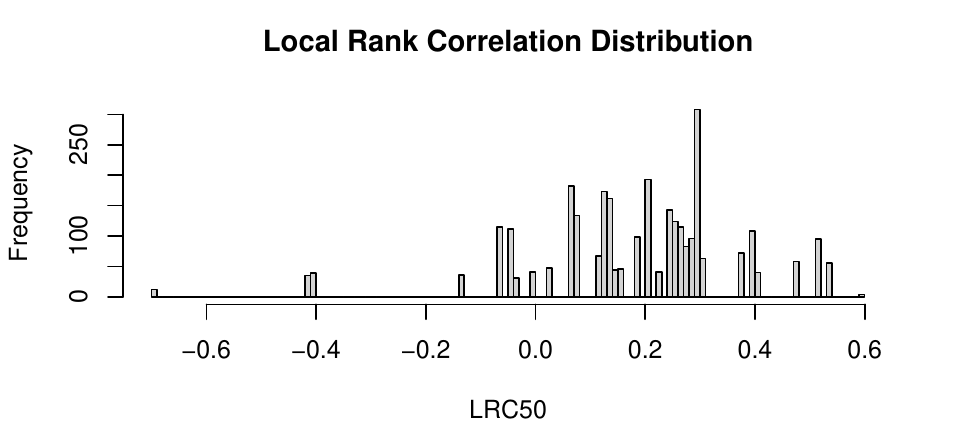}}
\caption{\label{Fig08} Here we see that the distribution of Local Rank Correlations (LRCs) within our 50 Clusters spans the range
from $-0.699$ to $+0.600$. Since this distribution contains only $50$ distinct numerical values, it certainly does not (and cannot)
look much like a ``continuous'' parametric distribution.} 
\end{figure}

Our choice of $K = 50$ clusters was determined within the \textit{Explore} phase when $Box-Whisker$ plots showed that the
distribution of estimated ``effect sizes'' would become excessively variable for $K > 50$. Specifically, any analyst applying
\textit{LC Strategy} would examine the ``LCcompare()'' graphic displayed in Figure~\ref{Fig06} and choose $K = 50$
for the very reasons outlined in the caption of that figure. Figure~\ref{Fig07} displays a histogram showing variation in
the sizes of these $50$ clusters. Finally, Figure~\ref{Fig08} shows that the distribution of LRC estimates looks somewhat
discrete simply because it contains only $50$ distinct numerical values.

\newpage

\section{LC ``Reveal'' Stage: randomForests, Partial Dependence Plots and a Single RP Tree} 

We can now stop ``Exploring'' and enter the (final) ``Reveal'' phase of \textit{NU Learning}. Specifically, we
now look across, rather than only within, our $K = 50$ clusters of relatively well-matched US
Counties that provide the most (potentially causal) \textit{Added Information} about relationships
between the AACRmort and Bvoc ``measures'' of mortality and air pollution, respectively. This new
information is embedded within the \textit{LRC50} variable that takes on only one of $50$ distinct numerical
values between $-0.699$ and $+0.600$ for each US County and is displayed in Figure~\ref{Fig08}.

\begin{figure}[!htbp]
\center{\includegraphics[width=4in]{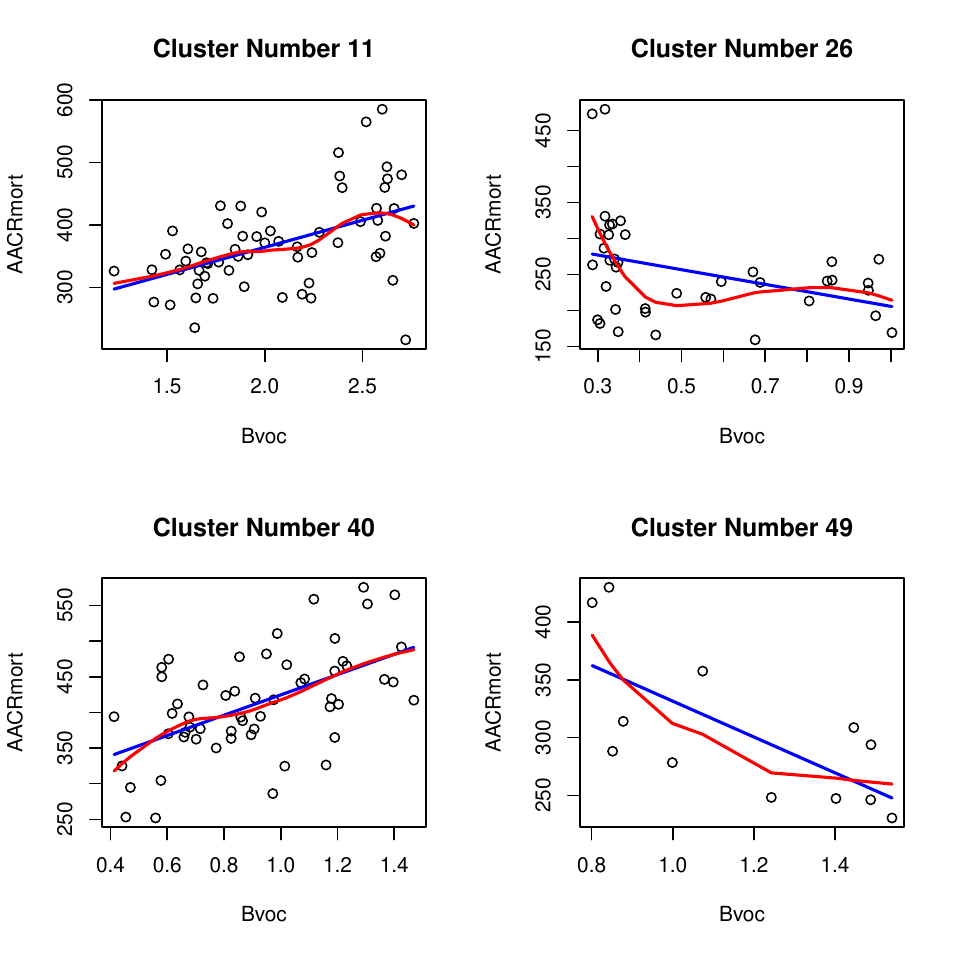}}
\caption{\label{Fig09} Here are plots of Bvoc versus AACRmort within four of the $50$ Clusters.
The straight blue lines are least-squares fits, while the red curves are ``loess'' smooths (span = $0.75$)
for predicting AACRmort from Bvoc. While the (Spearman) LRCs within the two left-hand Clusters
($11$ and $40$) are positive, both right-hand clusters ($26$ and $49$) are typical of the
negative LRC estimates observed for a total of $421$ US Counties ($21.3\%$ of the $2,973$ US
Counties without missing data).} 
\end{figure}

\subsection{Interaction Effects Abound}

Figure~\ref{Fig09} contains plots of Bvoc versus AACRmort for $4$ of our $50$ Clusters. The two left-hand clusters have positive
LRC-estimates and blue least-squares lines that slope upwards, while the two right-hand clusters have negative LRC-estimates and
blue least-squares lines that slope downwards. Actually, $421$ US Counties ($21.3\%$) of all $2,973$ US Counties without missing
data fall into the $10$ out of $50$ Clusters (i.e. $20\%$) that yield \textit{negative LRC-estimates}. In other words, almost
$80\%$ of US Counties are in the $80\%$ of Clusters with strictly positive LRC-estimates. After all, the Rank-Correlation for
the overall Bvoc vs. AACRmort scatter displayed in Figure~\ref{Fig01} is $+0.474$. This made us wonder whether larger clusters
might tend to produce positive LRC estimates? 

\begin{figure}[!htbp]
\center{\includegraphics[width=4in]{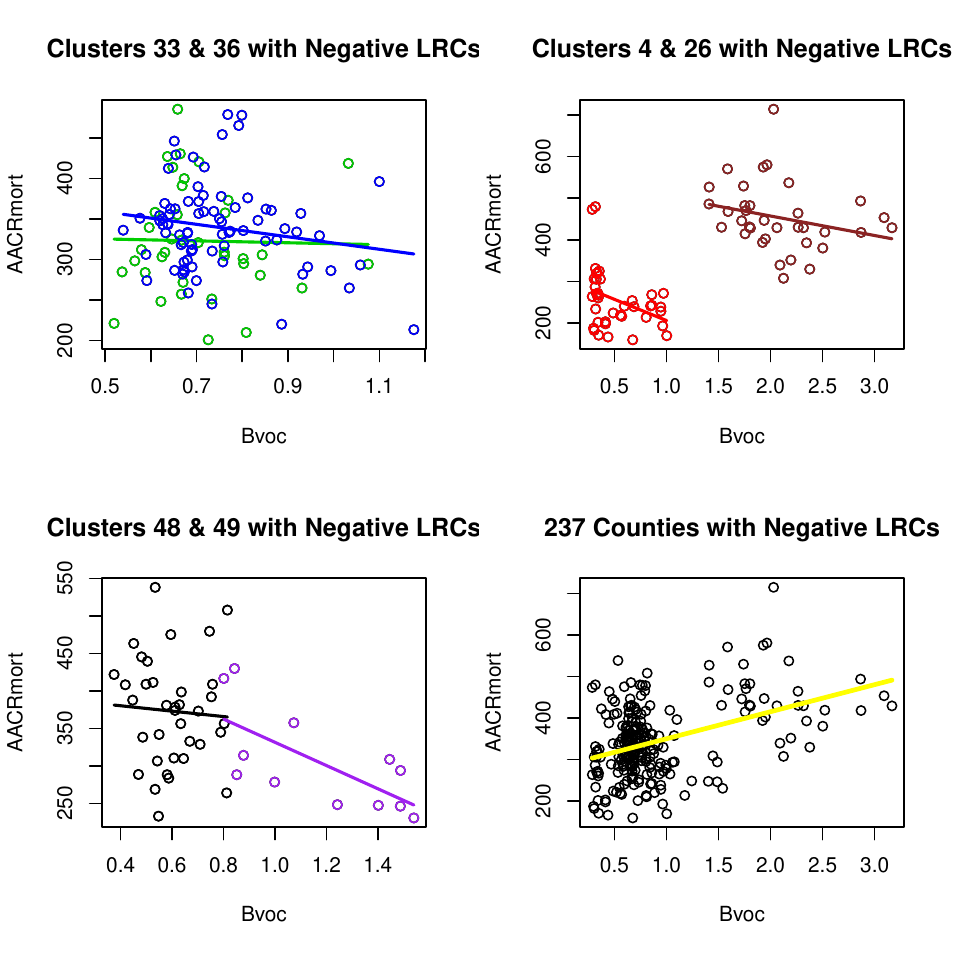}}
\caption{\label{Fig10} When AACRmort and Bvoc data from the 6 clusters with most Negative LRC estimates are merged together, all
other information about differences between these US Counties is discarded. In the lower-right panel, we see that their ``Merged
Trend'' becomes \textit{clearly positive} ...shown in Yellow. Each of the other three panels show AACRmort vs. Bvoc scatters for
only two of the six clusters with negative LRCs ...using Brown and Red, Blue and Green, or Purple and Black points and ``Local
Trend'' lines.}  
\end{figure}

An extreme illustration of this is displayed in the lower-right panel of Figure~\ref{Fig10}. Data from the $237$ Counties
within the $6$ clusters with the most Negative LRCs are merged together there, yielding a scatter with clearly positive slope (and
an LRC of $+0.267$.) Meanwhile, each of the other $3$ panels shows scatters for only $2$ out of these $6$ clusters with most negative
LRCs. Clearly, clustering of US Counties on their confounded $X-$characteristics predictive of AACRmort \textit{truly does matter}!

\subsection{Insights from a randomForest of 500 Trees}

Given a random forest of $500$ tree models for prediction of LRC estimates [Breiman (2001,2002), Liaw and Wiener (2002 - 2022)], the Partial Dependence Plot (PDP), Friedman (2001), for one potential predictor depicts the \textit{marginal relationship} that results from averaging over all other potential predictors. This marginal relationship ignores potential interaction effects and can be linear, monotonic or more complex.

\begin{figure}[!htbp]
\center{\includegraphics[width=4in]{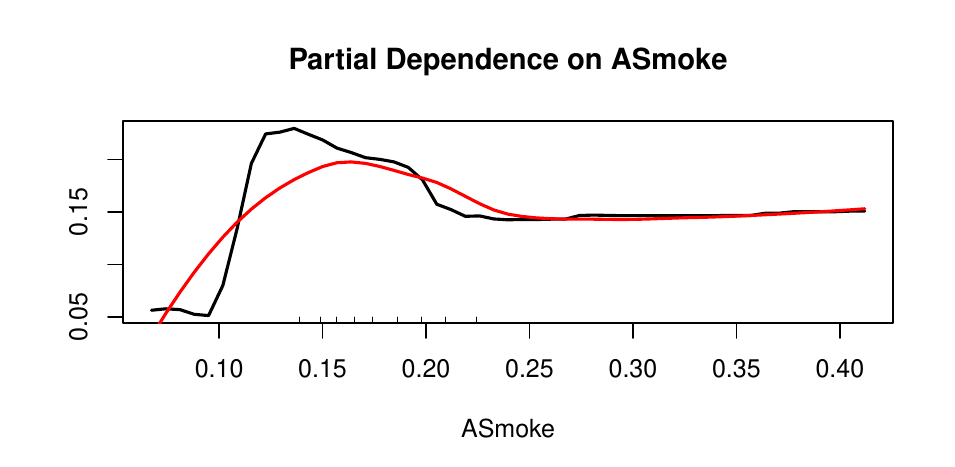}}
\caption{\label{Fig11} The single most important predictor of LRCs is the local Adult Smoking rate (ASmoke): $\%$IncMSE$=71.28$.
ASmoke rates within the $0.15$ to $0.20$ range are predictive of highly positive LRCs between AACRmort and Bvoc. This marginal
relationship is not monotone; ASmoke rates above $.25$ correspond to somewhat lower but still relatively high LRCs.}
\end{figure}

\begin{figure}[!htbp]
\center{\includegraphics[width=4in]{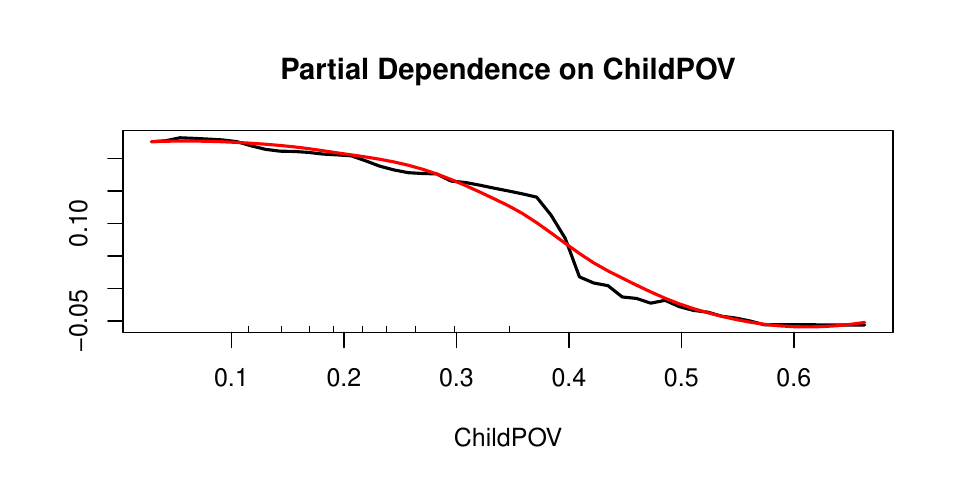}}
\caption{\label{Fig12} The second most important predictor of LRCs is ChildPOV: $\%$IncMSE$=59.46$.
When Child Poverty rates are less than $.30$, LRCs are strongly positive. However, LRCs monotonically decrease
and can even become negative when ChildPOV exceed $0.50$. In fact, the ChildPOV fate is the only predictor associated
with strictly negative LRC predictions.} 
\end{figure}

\begin{figure}[!htbp]
\center{\includegraphics[width=4in]{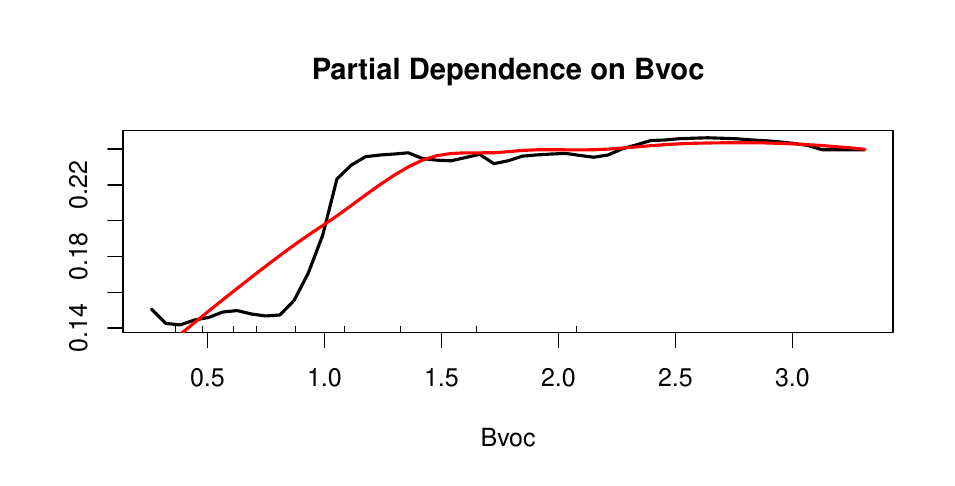}}
\caption{\label{Fig13} The third most important predictor of LRCs is PREMdeath: $\%$IncMSE$=56.31$.
PREMdeath is expressed as a count of ``premature deaths'' per $100,000$ county residents. As this rate
increases from $5,000$ to $15,000$, LRCs steadily increase from $0.15$ to about $0.28$. This marginal
relationship is rather close to being monotone increasing. Like ASmoke, PREMdeath appears to be
a key determinant of excess Circulatory and/or Respiratory mortality.} 
\end{figure}

\begin{figure}[!htbp]
\center{\includegraphics[width=4in]{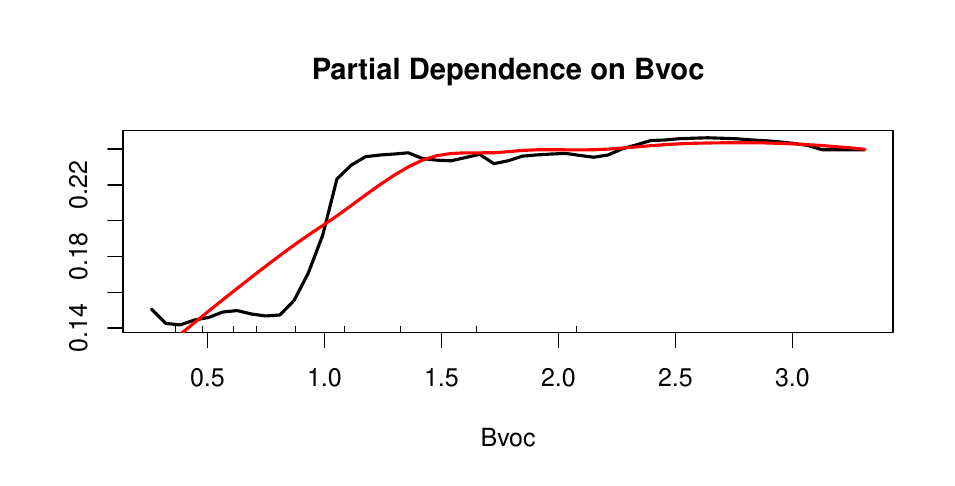}}
\caption{\label{Fig14} The fourth most important predictor of LRCs (Within-Cluster Spearman rank
correlations between AACRmort and Bvoc) is Bvoc itself: $\%$IncMSE$=46.40$. This marginal
relationship is also close to being monotone increasing. In fact, LRC's of approximately $0.24$
correspond to all Bvoc levels above $1.20$.}
\end{figure}

\begin{figure}[!htbp]
\center{\includegraphics[width=4in]{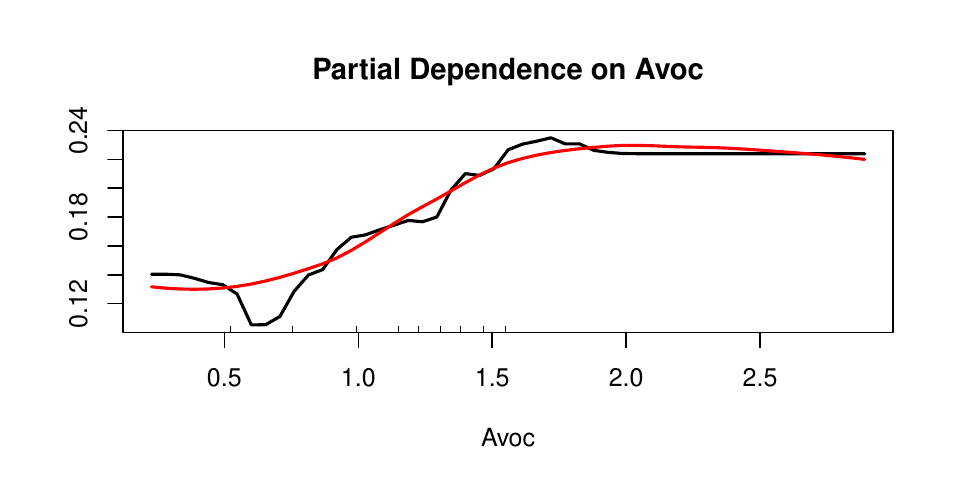}}
\caption{\label{Fig15} The predictor of LRCs in fifth place is $pmSO4: \%$IncMSE$=37.82$. This
plot suggests a complex marginal relationship: LRCs steadily increase over the range $0.4 \leq pmSO4 \leq 1.4$, 
but then drop sharply when $pmSO4$ exceeds $1.4$.}   
\end{figure}

\begin{figure}[!htbp]
\center{\includegraphics[width=4in]{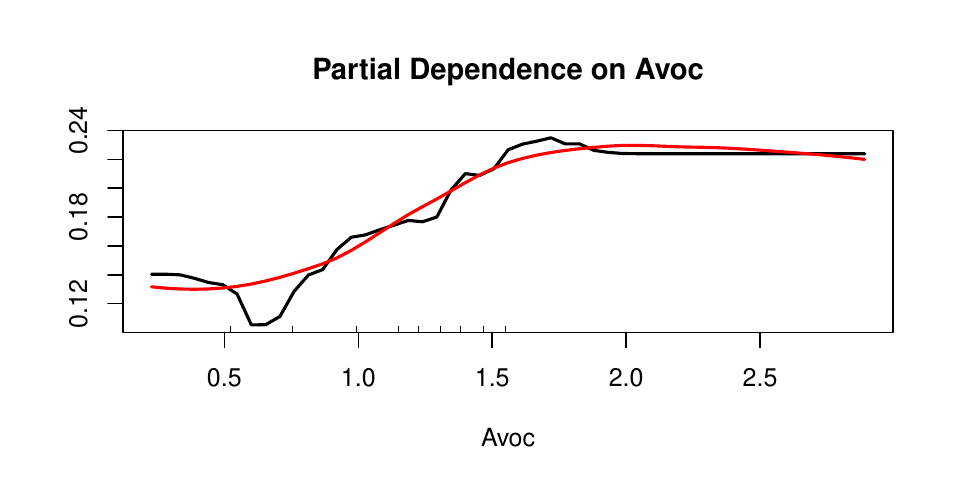}}
\caption{\label{Fig16} The sixth-place predictor of LRCs is Avoc: $\%$IncMSE$=25.26$. Here, LRC levels
vaguely approximate a monotonically increasing marginal relationship, except for a rather large ``dip''
near Avoc$= 0.6$ and a small decrease for Avoc $> 1.7$. The general shape of this marginal relationship
is somewhat like that of the Bvoc predictor in Figure~\ref{Fig14}.}   
\end{figure}

\begin{figure}[!htbp]
\center{\includegraphics[width=4in]{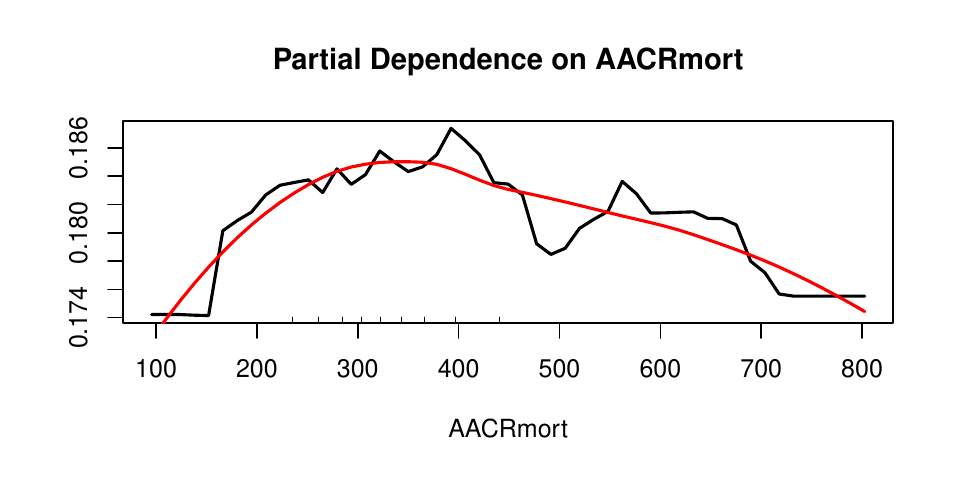}}
\caption{\label{Fig17} A seventh predictor of LRCs (between AACRmort and Bvoc) is AACRmort
itself: $\%$IncMSE$=17.79$. Since the vertical range in this plot is quite small ($0.014$),
any suggestion of ``curvature'' in this marginal relationship is probably unimportant.} 
\end{figure}

\begin{figure}[!htbp]
\center{\includegraphics[width=4in]{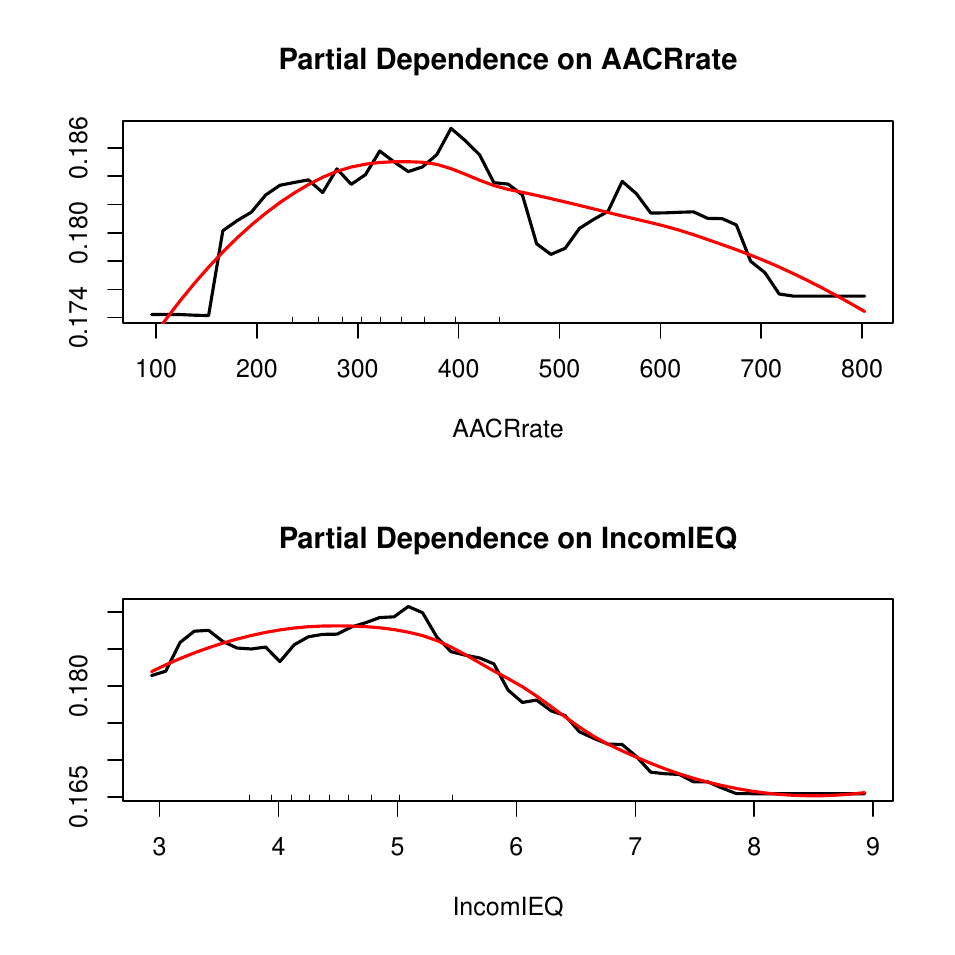}}
\caption{\label{Fig18} The last (and least important) predictor of within-cluster LRCs is IncomIEQ: 
$\%$IncMSE$=16.15$. This marginal relationship is complex but appears to monotonically decrease
after a local mode near IncomIEQ$ = 5$.}  
\end{figure}

The highly detailed captions for the next eight Figures (\ref{Fig11} through~\ref{Fig18}) outline our interpretations of these marginal
relationships. 

Table 3 summarizes the key characteristics of our eight PDP plots. Note that the top 4 $X-$confounder 
variables predictive of AACRmort include the three variables inducing highest Incremental Node Purity.

\begin{table}[h]
\begin{tabular}{@{}ccc@{}}
\multicolumn{3}{c}{\textbf{TABLE 3 -- Importance Statistics for $8$ Variables}} \\
\textbf{Variable}  &  \textbf{\%IncMSE}   &  \textbf{IncNodePurity} \\
ASmoke    &  71.28070   &    16.611216	\\	   
ChildPOV  &  59.45606   &    17.160771	\\  
PREMdeath &  56.31122   &    11.765453	\\
Bvoc      &  46.40073   &    16.198999	\\
pmSO4     &  37.81722   &    12.071959	\\
Avoc      &  25.25779   &    14.658269	\\   
AACRmort  &  17.79182   &     4.878167	\\
IncomIEQ  &  16.14760   &     5.173446 \\
\end{tabular}
\end{table}

\subsection{Insights from a Single Recursive Partitioning Tree}

\begin{figure}[!htbp]
\center{\includegraphics[width=6.5in]{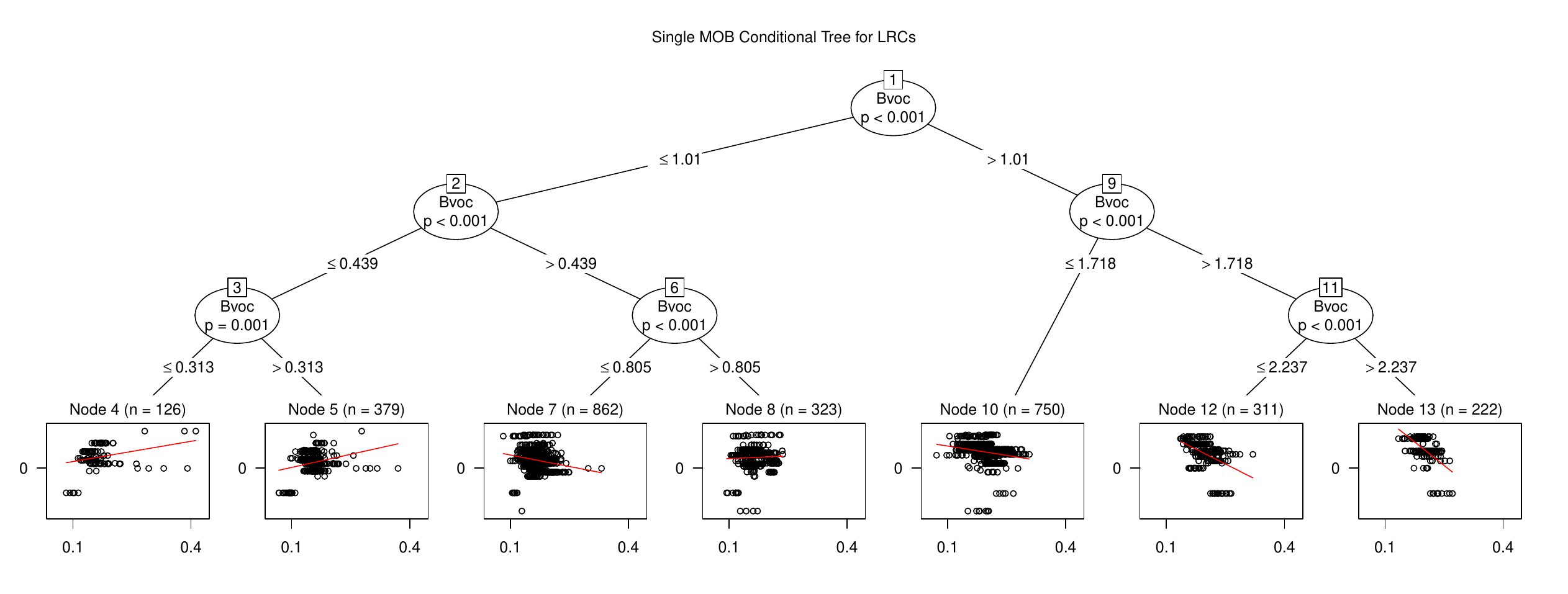}}
\caption{\label{Fig19} All 7 plots in the bottom row of this tree show $ASmoke$ rates that vary from less than 0.1 (i.e. $10\%$ of residents) to more than 0.4 ($40\%$) on their horizontal axis. On each vertical axis, estimated $LRC$s range from $-0.7$ to $+0.6$.
Note that this RP Tree suggests that high ASmoke rates are particularly harmful within the two left-hand plots corresponding to low
Bvoc $< 0.44$ intervals. Within four of the five remaining higher Bvoc intervals on the the right-hand side of this MOB (model-based)
tree, LRCs (between AACRmort and Bvoc) surprisingly appear to actually \textit{decrease} as local ASmoke rates \textit{increase}!}  
\end{figure}

We used the ``model based'' algorithms of Zeileis, Hothorn and Hornik (2008, 2015, 2022) to develop the relatively simple
conditional regression-tree displayed in Figure~\ref{Fig19}. This tree has a maximal depth of $3$ and defines $7$ optimal ``Bvoc intervals''. Within each interval, observed LRCs are predicted from local ASmoke levels using simple linear regression. Summary
statistics for this tree are listed in Table $4$.

\begin{table}[ht]
\begin{tabular}{@{}cccccc@{}}
\multicolumn{6}{c} {\textbf{TABLE 4 -- Statistics for LRC Prediction within 7 Bvoc Ranges}} \\
Final & Bvoc & Bvoc & LRC & ASmoke & US \\
Node & Range & Level & Intercept & Slope & Counties \\
4 & 0.261-0.313 & Low & -0.00411 & +1.09902 & 126 \\
5 & 0.314-0.439 & Low & -0.13183 & +1.42106 & 379 \\
7 & 0.440-0.805 & Medium & 0.33623 & -1.24350 & 862 \\
8 & 0.806-1.010 & Medium & 0.12447 & +0.29020 & 323 \\
10 & 1.011-1.718 & High & 0.45914 & -1.01360 & 750 \\
12 & 1.719-2.237 & High & 0.86823 & -3.21152 & 311 \\
13 & 2.238-3.309 & High & 1.31711 & -5.09834 & 222 \\     
\end{tabular}
\end{table}
\vspace{0.5cm}

\subsection{Advantages of Local Comparisons}

Personal computers have helped shape statistical theory as well as its practice over the last 50 years. Freely available software
can provide computational and visual fast-tracks into the strengths and weaknesses of alternative statistical methods. For example,
the lower right-hand panel of Figure~\ref{Fig10} showed just how misleading a scatter-plot can be that ignores cluster-membership.
Our clusters contain US Counties with most similar $X-$characteristics, so they compare only ``apples'' with other ``apples'',
``potatoes'' with other ``potatoes'', etc. 

Our LC strategy leads to \textit{fair local comparisons} when the exposure variable is binary and to cogent
\textit{local inferences} when that exposure (say, Bvoc) is both continuous and correlated with potential confounding variables
(such as Avoc, ASmoke, etc.)

A key feature of LC Strategy is that it creates a ``new'' variable of genuine interest for all observational units within each
cluster. This variable contains either \textit{Local Treatment Differences} (LTDs) when the exposure variable is Binary
(New Treatment vs Standard Treatment) [Lopiano, Obenchain and Young (2014)] or else (Spearman) \textit{Local Rank Correlations}
(LRCs) when the exposure varies continuously. In this paper, our estimated LRCs quantify important new insights into potential
relationships among environmental measures of genuine current interest.

\section{Summary}   

Our eight PDP plots in Figures (\ref{Fig11} through~\ref{Fig18}) and the summary statistics from Table 3 have
encouraged us to focus our attention primarily on the \textit{top-four predictors of LRCs}, where $Bvoc$ is number four. Again,
neither $pmSOA \equiv Avoc + Bvoc$ nor either of its two components can apparently be actually ``measured'' today ...using existing
scientific instruments with reasonable accuracy.

Bvocs include \textit{terpenes} from trees and grass, and EPA Bvoc \textit{predictions} are, at least on average, positively
correlated with corresponding local CDC CRmort counts, before and after being Age Adjusted. How will individual US environmentalists
react to this ``news''? Luckily, our single RP MOB tree in Figure~\ref{Fig19} does suggest that levels of Bvoc $> 1.01$ could be
protective against high AACRmort rates when ASmoke rates are either average or higher!

\begin{figure}[!htbp]
\center{\includegraphics[width=6.5in]{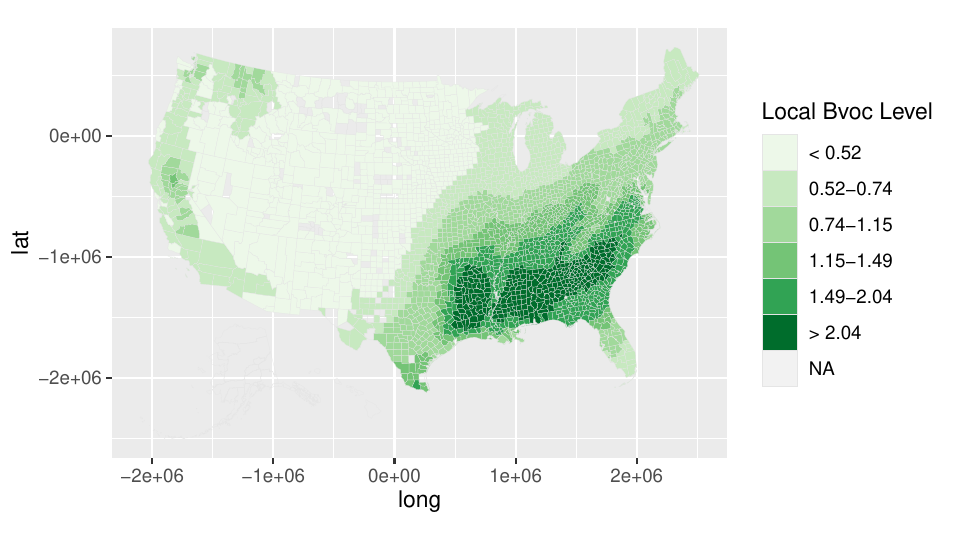}}
\caption{\label{Fig20} This US map [Kahle and Wickham (2013), Wickham (2016)] shows variation in EPA Bvoc predictions across US
Counties using six shades of Green. Low Bvoc levels receiving the lightest shade of Green dominate the US ``Great Plains'' and Rocky Mountain regions where agricultural production is either low or supported by irrigation. Note that a vast majority of counties receive
the same Bvoc color as their adjacent counties. Note also that 151 US Counties fall into the $7^{th}$ ``NA'' category and
receive the see-through ``background'' color.}  
\end{figure}

\begin{figure}[!htbp]
\center{\includegraphics[width=6.5in]{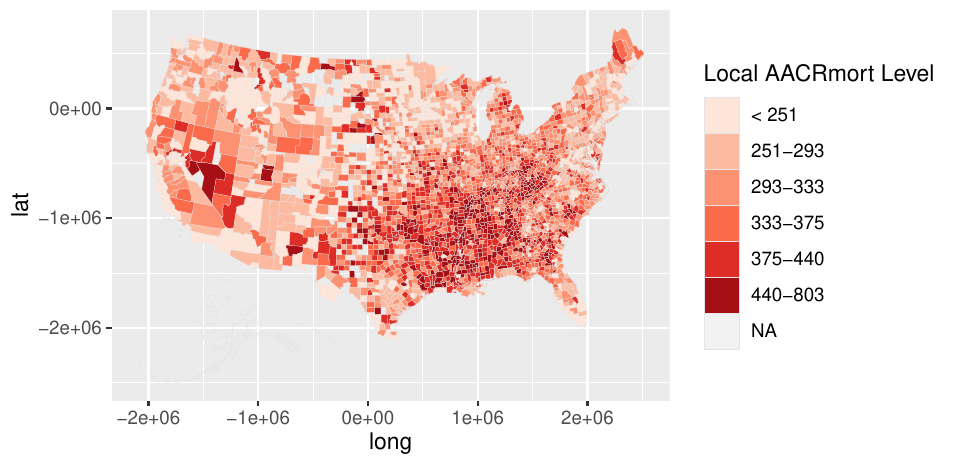}}
\caption{\label{Fig21} This map shows variation in AACRmort counts per $100,000$ residents using six shades of Red. These adjusted
counts are certainly not smooth functions of geographic latitude and longitude; variation in AACRmort counts may result primarily
from variation in the socioeconomic characteristics of local residents. After all, the vast majority of adjacent counties receive a
somewhat different or very different shade of Red.}  
\end{figure}

\begin{figure}[!htbp]
\center{\includegraphics[width=6.5in]{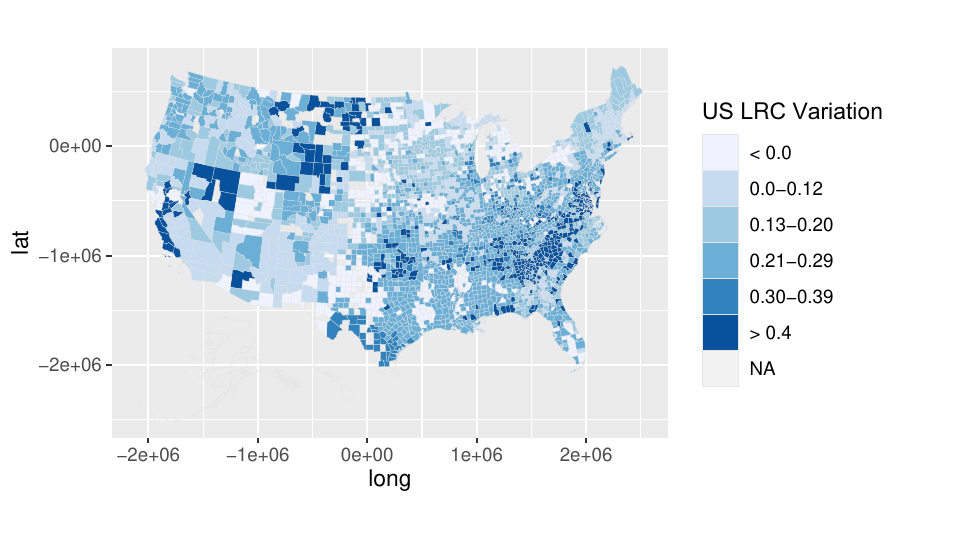}}
\caption{\label{Fig22} On this map showing variation in \textit{Local Rank Correlations (LRCs)} across US Counties, six shades
of Blue identify LRC levels of association between AACRmort and Bvoc within our 50 clusters of US counties well-matched in the
$7-$dimensional space of potential confounded characteristics. Note that all adjacent counties receiving different shades of Blue
must be in different Clusters and thus have somewhat different characteristics. All $421$ counties with negative LRC estimates
receive the lightest shade of Blue. Again, 151 counties without data receive the see-through ``background'' color.}  
\end{figure}

Our final three Figures (\ref{Fig20},~\ref{Fig21} and~\ref{Fig22}) display US Maps that use shades of green, red or blue to color individual US Counties and, thereby, reveal interesting differences in patterns of variation in Bvoc, AACRmort or LRCs, respectively.
These graphics illustrate just how varied and/or different these three quantities apparently are.  

The future of \textit{NU Learning} approaches for analysis of cross-sectional observational data strikes us as being quite promising.
Unfortunately, traditional (Parametric and Supervised Learning) models used in Propensity Score estimation tend to be global
(rather than local) and may make strong and potentially unrealistic assumptions.

Most interesting questions don't have an answer that is clearly ``Black'' or ``White''. Data from well-designed and well-executed experiments can illuminate some key issues. Meanwhile, large collections of observational data can reveal interesting real-world ``shades of gray''.

Our use of LC Strategy in our analysis of rather voluminous and detailed data highlights a truly key advantage of our
computer intensive approach. Namely, this approach generates many \textit{highly informative graphical displays} that
range from rather simplistic to quite detailed. Examining a wide variety of graphics can guide further analyses in unanticipated
directions. For example, note the great detail embedded within Figures \ref{Fig19},~\ref{Fig20},~\ref{Fig21} and~\ref{Fig22}.
Such plots enable comparisons among US counties that are ``similar'' in terms of either \textit{Cluster Membership} in three dimensions (AACRmort, Bvoc and ASmoke) or \textit{Geography} in two dimensions (latitude and longitude).

\subsection{Contributions to Data and Code Sharing}

Our reconstruction of the data used in Pye et al. (2021) can be found in the \textit{pmdata} data.frame that is part of
the newest Version ($1.4$) of the \textit{LocalControlStrategy} R-package, Obenchain (2015-2022). Alternatively, a CSV file
containing only $25$ key variables can be downloaded from \textit{dryad}, Young and Obenchain (2022),
\url{https://doi.org/10.5061/dryad.63xsj3v58}. These data are as much ``like'' the CDC and EPA data used by Pye et al. (2021)
as is possible by private US citizens.

Internet site \url{http://localcontrolstatistics.org} contains an archive, \textit{LCS$\_$EPA$\_$archive.zip}, that can be freely
downloaded. This archive contains $R-$code that performs the basic analyses and displays examples of figures from this paper. A
wide variety of introductory information about using \textit{LC Strategy} for statistical \textit{NU Learning} can also be read at
or downloaded from this site.

\subsection{Insights from a New Statistical Perspective}

Our use of \textit{LC Strategy} has lead to findings that are fundamentally different from and quite possibly more relevant
and technically sound than those of Pye et al. (2021). Since interaction effects abound within the data, it's certainly not surprising
that our LRC distribution (Figure~\ref{Fig08}) from simple local (within cluster) models and our $RP-$Tree model of Figure~\ref{Fig19}
each provide insights more clear and informative than complex EPA models. While EPA models can be claimed to be ``smooth and continuous'',
they are embedded within an abstract \textit{high-dimensional} $X-$space of confounded variables.

\subsection{Our Focus on Biogenic Volatile Organic Compounds}

It became clear in our earliest data analyses that $Bvoc$ would be a better predictor of $AACRmort$ than either
$Avoc$ or $pmSOA \equiv Avoc + Bvoc$. In fact, as shown in Table 5, the Spearman Rank Correlations between $Bvoc$ and
all six other key measures are higher than those of $Avoc$. Only the single Pearson correlation of $Avoc$ with Sulfates
($pmSO4$) is larger than its corresponding correlation with $Bvoc$. Furthermore, all $\beta-$coefficient estimates of
$Avoc$ effects on $AACRmort$ are \textit{negative or zero} in Figure~\ref{Fig02}.

\begin{table}[ht]
\begin{tabular}{@{}lccccccc@{}}
\multicolumn{8}{c}{\textbf{TABLE 5 -- Pearson and Spearman  Correlations between Variables}} \\
Method   &      & AACRmort & PREMdeath & pmSO4 & ASmoke & ChildPOV & IncomIEQ \\  
-------  & ---- & ------ & ------ & ------ & ------ & ------ & ------ \\
Pearson  & Bvoc & 0.4589 & 0.4217 & 0.5023 & 0.4622 & 0.4884 & 0.4163 \\  
Pearson  & Avoc & 0.2489 & 0.0890 & 0.7228 & 0.3271 & 0.1134 & 0.1393 \\  
-------  & ---- & ------ & ------ & ------ & ------ & ------ & ------ \\
Spearman & Bvoc & 0.4740 & 0.4808 & 0.6195 & 0.5437 & 0.4950 & 0.4154 \\  
Spearman & Avoc & 0.2458 & 0.1475 & 0.6137 & 0.3763 & 0.1145 & 0.1467 \\  
\end{tabular}
\end{table}

\subsection{Revealing Effect Heterogeneity}

As is clear from Figure~\ref{Fig08}, our distribution of LRC estimates resulting from using $K = 50$ ``Ward.D'' Clusters
displays \textit{considerable heterogeneity}. Furthermore, $40$ of our $50$ Clusters yield positive LRC estimates between $AACRmort$
and $Bvoc$ and contain almost $86\%$ of the $2,973$ US Counties without missing data. Table $6$ displays some added detail on
the $16$ Clusters with ``highly significant'' (one-tailed) probabilities. Only the $3$ Clusters listed at the bottom of Table $6$ have
negative LRC estimates and contain a total of $86$ US Counties; these are the only $86$ US locations where AACRmort levels tend to monotonically decrease as Bvoc levels increase. The genuinely ``bad news'' here is that $6$ of the $13$ Clusters at the top of Table $6$ have positive LRCs that are even more highly significant than the most significant negative LRCs! Unfortunately, within these $916$ US
Counties, AACRmort levels tend to monotonically increase as Bvoc levels increase! Finally, we note that $Bvoc = terpenes + isoprenes$, where terpenes clearly appear to be the more ``active'' sub-component of Bvoc. Data on pmTOT ($PM_{2.5}$) and $27$ of its sub-components are contained in our \textit{pmdata} \textbf{R} data.frame.
 
\vspace{0.5cm}
\begin{table}[ht]
\begin{tabular}{@{}ccccc@{}}
\multicolumn{5}{c} {\textbf{TABLE 6 -- 16 Clusters with Highly Significant (Spearman) LRCs}} \\
Cluster & Number of    &  Spearman & One-Tailed  & Sig-Level \\
Number  & US Counties  &    LRC    & Probability &   Symbol  \\
 1  &  83  & +0.277 & 0.006   & $**$ \\
 2  &  52  & +0.393 & 0.002   & $**$ \\
 6  &  58  & +0.478 & 0.0001  & $****$ \\ 
 8  &  63  & +0.303 & 0.008   & $**$ \\
10  &  56  & +0.391 & 0.002   & $**$ \\
11  &  61  & +0.511 & 0.00002 & $****$ \\
14  & 115  & +0.267 & 0.002   & $**$ \\ 
17  &  34  & +0.518 & 0.001   & $***$ \\
32  & 127  & +0.292 & 0.0005  & $***$ \\
39  &  72  & +0.376 & 0.0006  & $***$ \\
40  &  56  & +0.537 & 0.00002 & $****$ \\
41  &  99  & +0.295 & 0.002   & $**$ \\
46  &  40  & +0.405 & 0.005   & $**$ \\  
--  &  --  & ------	& ------	  & ----- \\											   
 4  &  35  & -0.414 & 0.007   & $**$ \\ 
26  &  39  & -0.404 & 0.006   & $**$ \\
49  &  12  & -0.699 & 0.007   & $**$ \\
\end{tabular}
\end{table}
\vspace{0.5cm}

\subsection{Our Bottom Line}

There appear to be only two realistic conclusions from the EPA data that we analyze here. Either biogenic
volatile organic compounds are real \textit{killers}, or current EPA models for the chemical content of air
pollution are misleadingly wrong.
 
\subsection*{Conflict of Interest}

As independent and self-funded researchers, the authors declare that no competing interests exist.

\end{document}